\documentclass[reprint,amsmath,amssymb,aps,prb,showpacs,floatfix]{revtex4-1}

\usepackage[dvips]{color,graphicx}
\DeclareGraphicsExtensions{.eps, .jpg, .png}
\usepackage{dcolumn}

\begin{document}

\title{Structures and stability of calcium and magnesium carbonates at
  mantle pressures}

\author{Chris J.\ Pickard} \email{c.pickard@ucl.ac.uk}
\affiliation{Department of Physics \& Astronomy, University College
  London, Gower Street, London WC1E~6BT, UK\\
  London Institute for Mathematical Sciences, 35a South Street,
  Mayfair, London, W1K 2XF, UK}

\author{Richard J.\ Needs} \affiliation{Theory of Condensed Matter
  Group, Cavendish Laboratory, J J Thomson Avenue, Cambridge CB3 0HE,
  UK}

\date{\today}

\begin{abstract}

  \textit{Ab initio} random structure searching (AIRSS) and density
  functional theory methods are used to predict structures of calcium
  and magnesium carbonate (CaCO$_3$ and MgCO$_3$) at high pressures.
  We find a previously unknown CaCO$_3$ structure which is more stable
  than the aragonite and ``post aragonite'' phases in the range 32--48
  GPa.  At pressures from 67 GPa to well over 100 GPa the most stable
  phase is a previously unknown CaCO$_3$ structure of the pyroxene
  type with fourfold coordinated carbon atoms.  We also predict a
  stable structure of MgCO$_3$ in the range 85--101 GPa.  Our results
  lead to a revision of the phase diagram of CaCO$_3$ over more than
  half the pressure range encountered within the Earth's mantle, and
  smaller changes to the phase diagram of MgCO$_3$.  We predict
  CaCO$_3$ to be more stable than MgCO$_3$ in the Earth's mantle above
  100 GPa, and that CO$_2$ is not a thermodynamically stable compound
  under deep mantle conditions.  Our results have significant
  implications for understanding the Earth's deep carbon cycle.

\end{abstract}

\pacs{64.70.K-, 71.15.Mb, 61.50.Ks, 62.50.-p}


\maketitle




\section{Introduction}

The occurrence of CO$_2$ within magmas and volcanic gases indicates a
significant carbon presence within the Earth's lower mantle
\cite{Marty_1987,Halliday_2013}.  Carbon has a low solubility in
mantle silicates and the majority of the oxidized carbon in the
Earth's mantle is believed to exist in the form of carbonates.
Calcium and magnesium carbonate (CaCO$_3$ and MgCO$_3$) are the main
sources and sinks of atmospheric CO$_2$ within the Earth's mantle.
Carbonates are conveyed into the deep Earth by subduction, and carbon
is recycled to the surface via volcanic processes in the form of
CO$_2$-containing fluids and solids, and diamonds
\cite{Ghosh_2009,Frezzotti_2011}.  However, the details of carbon
storage within the Earth's interior are unclear.  The Deep Carbon
Observatory \cite{DCO} has been set up to investigate carbon within
the Earth's deep interior.  CaCO$_3$ and MgCO$_3$ play fundamental
roles in the global carbon cycle and influence the climate of our
planet \cite{Dasgupta_2010,Hazen_2013_DCO}.  Knowledge of the
structures, energetics and other properties of CaCO$_3$ and MgCO$_3$
at high pressures is therefore important in understanding the Earth's
mantle, and especially the carbon cycle.

The low-pressure calcite form \cite{Bragg_structure_of_calcite_1914}
of CaCO$_3$ is one of the most abundant minerals on the Earth's
surface and is the main constituent of metamorphic marbles.  Several
metastable calcite-like phases have been observed
\cite{Bridgman_1939,Suito_2001,Merlini_2012}, and a calcite-related
phase has been reported at around 25 GPa
\cite{Catalli_2005,Merlini_2012}.  At pressures of about 2 GPa calcite
transforms to the aragonite structure
\cite{Bragg_structure_of_aragonite_1924} of $Pnma$ symmetry.  At about
40 GPa aragonite transforms into the ``post aragonite'' ($Pmmn$)
structure of CaCO$_3$, which is stable up to at least 86 GPa
\cite{Ono_2005,Oganov_2006}.  The low pressure magnesite phase of
MgCO$_3$ has the same structure as calcite.  Experiments indicate that
magnesite is stable up to 80 GPa \cite{Fiquet_2002}, and a phase
transition occurs above 100 GPa to an unknown magnesite II structure
\cite{Isshiki_2004_magnesite_and_high_pressure_form,Boulard_2011}.

\section{Structure searches}

Density functional theory (DFT) calculations for high pressure phases
of CaCO$_3$ and MgCO$_3$ were performed by Oganov \textit{et al.}\
using an evolutionary structure searching algorithm
\cite{Oganov_2006,Oganov_2008}.  These calculations predicted a
transition from the calcite to aragonite to ``post aragonite''
structures of CaCO$_3$, followed by a transition to a structure of
$C222_1$ symmetry at pressures over 100 GPa.  Similar calculations for
MgCO$_3$ predicted transitions from magnesite to a structure of $C2/m$
symmetry at 82 GPa, followed by a transition to a structure of $P2_1$
symmetry at 138 GPa, and a phase of $Pna2_1$ symmetry at 160 GPa
\cite{Oganov_2008}.

Calculations using the \textit{ab initio} random structure searching
(AIRSS) technique \cite{Airss_review} have led to the discovery of
structures that have subsequently been verified by experiment, for
example, in silane \cite{pickard_silane}, aluminium hydride
\cite{pickard_aluminum_hydride}, ammonia monohydrate
\cite{fortes_ammonia_monohydrate_II} and ammonia dihydrate
\cite{griffiths_ammonia_dihydrate_II}.  In the basic AIRSS approach a
cell volume and shape is selected at random from within reasonable
ranges, the atoms are added at random positions, and the system is
relaxed until the forces on the atoms are negligible and the pressure
takes the required value.  This procedure is repeated many times,
leading to a reasonably unbiased scheme which allows a significant
portion of the ``structure space'' to be investigated, although the
sampling may be rather sparse.  This approach is often successful for
small systems, but it involves sampling a large portion of the
high-energy structure space which is not normally of interest.  We
therefore reduce the size of the structure space investigated by
constraining the searches.

We first perform searches in small cells, constraining the initial
structures so that all of the atoms are at least 1 \AA\ apart.  The
low-enthalpy structures obtained from these calculations give us
information about the favorable bonding configurations and likely
nearest neighbor distances between the different atomic types.  At low
pressures we find that the low-enthalpy structures contain
well-defined triangular CO$_3$ or ring C$_3$O$_9$ units, and therefore
we place these units and Ca or Mg atoms randomly within the cells of
random shapes.  We ensure that the atoms are not too close together by
constraining the initial values of the minimum distances between atoms
for each of the six possible pairs of atomic species.  The six minimum
distances are obtained from low-enthalpy structures found in the
small-cell searches.  To construct the initial structures at higher
pressures we use minimum distances from low-enthalpy small-cell
structures to prepare new larger structures that approximately satisfy
the minimum distance constraints.  This approach helps to space out
the different species appropriately, while retaining a high degree of
randomness.  We perform searches at both low and high pressures, using
structures which are constrained to have a certain symmetry which is
enforced during the relaxation, but are otherwise random
\cite{Airss_review}.  This approach is useful because low energy
structures often possess symmetry \cite{Pauling_1929,Wales_1998},
although symmetry constraints break up the allowed structure space
into disconnected regions and can prevent some structures from
relaxing to lower energy ones \cite{Airss_review}.  We consider
structures containing up to eight formula units (f.u.) for CaCO$_3$
and twelve f.u.\ for MgCO$_3$.

Our first-principles DFT calculations are performed using the
\textsc{Castep} plane-wave basis set pseudopotential code
\cite{ClarkSPHPRP05}.  We use the Perdew-Burke-Ernzerhof (PBE)
generalized gradient approximation (GGA) density functional
\cite{Perdew_1996_PBE}, default \textsc{Castep} ultrasoft
pseudopotentials \cite{Vanderbilt90}, and a plane-wave basis set
energy cutoff of 440 eV.  We use a Brillouin zone sampling grid of
spacing $2\pi\times$0.1~\AA$^{-1}$ for the searches, and a finer
spacing of $2\pi\times$0.05~\AA$^{-1}$ for the final results reported
in this paper.

\section{C\lowercase{a}CO$_3$, pressure $\leq$ 50 GPa}

\begin{figure}
  \centering
  \includegraphics[width=0.4\textwidth,clip]{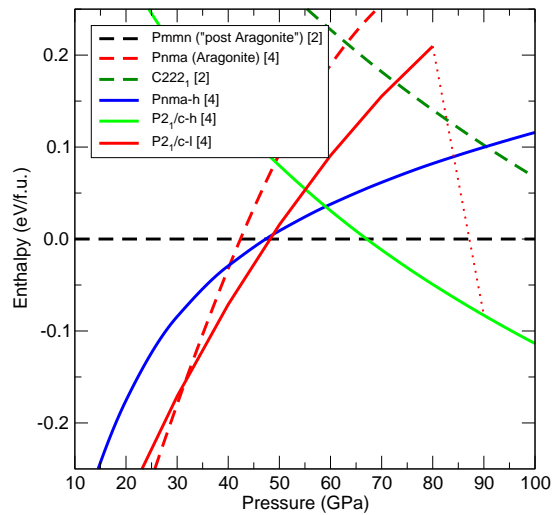}
  \caption{(Color online) Enthalpies per f.u.\ of CaCO$_3$ phases
    relative to ``post aragonite'', with the number of f.u.\ per
    primitive unit cell given within square brackets. The enthalpies
    of phases known prior to the current study are shown as dashed
    lines, while those found in the current study are shown as solid
    lines.  The dotted red line shows the collapse of the
    {CaCO$_3$-$P2_1/c$-$l$} structure into the more stable
    {CaCO$_3$-$P2_1/c$-$h$} structure at 80--90 GPa.}
  \label{fig:enthalpy_CaCO3}
\end{figure}

Calculated enthalpy-pressure curves for CaCO$_3$ phases are shown in
Fig.\ \ref{fig:enthalpy_CaCO3}, relative to the enthalpy of the ``post
aragonite'' phase.  The transition from aragonite to ``post
aragonite'' becomes energetically favorable at about 42 GPa, in
agreement with previous DFT results
\cite{Oganov_2006,Arapan_2007,Oganov_2008,Arapan_2010} and experiment
\cite{Ono_2005}.  We performed calculations for the CaCO$_3$-VI
structure reported in Ref.\ \onlinecite{Merlini_2012}, which was
suggested as a possible high pressure phase of CaCO$_3$.  However, we
found it to be very high in enthalpy, with a strongly anisotropic
stress and large forces on the atoms.  Relaxation of the CaCO$_3$-VI
structure at 40 GPa led to a reasonably stable structure with an
enthalpy close to that of aragonite, but the relaxed structure does
not have a region of stability on our phase diagram (Fig.\
\ref{fig:enthalpy_CaCO3}).  We also found a structure of $Pnma$
symmetry (``{CaCO$_3$-$Pnma$-$h$}'', where $h$ denotes ``high
pressure'') that is predicted to be more stable than aragonite above
40 GPa, and more stable than ``post aragonite'' below 47 GPa.
However, {CaCO$_3$-$Pnma$-$h$} does not have a region of thermodynamic
stability on our phase diagram because we find a previously unknown
structure of $P2_1/c$ symmetry (``{CaCO$_3$-$P2_1/c$-$l$}'', where $l$
denotes ``low pressure'') which is calculated to be the most stable
phase in the pressure range 32--48 GPa, see Fig.\
\ref{fig:enthalpy_CaCO3}.

At 42 GPa {CaCO$_3$-$P2_1/c$-$l$} is calculated to be about 0.05 eV
per f.u.\ more stable than aragonite and ``post aragonite'' and,
because these $sp^2$ bonded structures are similar, we expect that DFT
calculations should give rather accurate enthalpy differences between
them.  However, our {CaCO$_3$-$P2_1/c$-$l$} and {CaCO$_3$-$Pnma$-$h$}
structures do not provide as good a fit to the experimental X-ray
diffraction data as the ``post aragonite'' phase \cite{Oganov_2006}.
It is possible that large energy barriers hinder formation of the
{CaCO$_3$-$P2_1/c$-$l$} structure.  Another possibility is that the
laser-heated sample melts and the least stable polymorph crystallizes
from the melt first, in analogy to ``Ostwald's rule''
\cite{Ostwald_1897}.
In any case, the conditions within the Earth's mantle are not the same
as in diamond anvil cell experiments, and the timescales associated
with geological processes are enormously longer than those for
laboratory experiments.

\section{C\lowercase{a}CO$_3$, pressure $>$ 50 GPa}

At higher pressures we find another CaCO$_3$ structure of $P2_1/c$
symmetry (``{CaCO$_3$-$P2_1/c$-$h$}'') to be stable from 67 GPa to
well above 100 GPa.  Our {CaCO$_3$-$P2_1/c$-$h$} structure is about
0.18 eV per f.u.\ more stable than the $C222_1$ structure found by
Oganov \textit{et al.}\ \cite{Oganov_2006}, see Fig.\
\ref{fig:enthalpy_CaCO3}, and $C222_1$ does not have a region of
thermodynamic stability.  We also find that at about 80--90 GPa
{CaCO$_3$-$P2_1/c$-$l$} transforms into the more stable
{CaCO$_3$-$P2_1/c$-$h$} structure without any apparent energy barrier
(dotted red line in Fig.\ \ref{fig:enthalpy_CaCO3}).  Our calculations
lead to the prediction of a new and more stable polymorph of CaCO$_3$
at pressures $>67$ GPa.

\section{M\lowercase{g}CO$_3$}

\begin{figure}
  \centering
  \includegraphics[width=0.4\textwidth,clip]{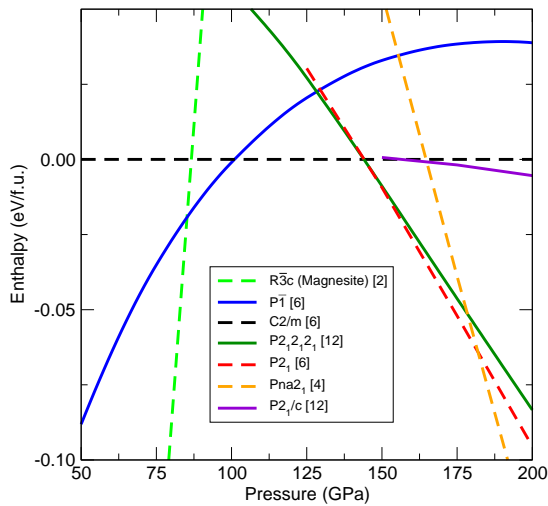}
  \caption{(Color online) Enthalpies per f.u.\ of MgCO$_3$ phases
    relative to the $C2/m$ phase, with the number of f.u.\ per
    primitive unit cell given within square brackets.  Previously
    known phases are shown as dashed lines, and those found in the
    current study are shown as solid lines.}
  \label{fig:Enthalpies_MgCO3_higher_pressure}
\end{figure}

Calculated enthalpy-pressure curves for MgCO$_3$ phases in the
pressure range 50--200 GPa are shown in Fig.\
\ref{fig:Enthalpies_MgCO3_higher_pressure}, relative to the $C2/m$
phase.  We find a previously unreported structure of $P\bar{1}$
symmetry to be the most stable in the range 85--101 GPa.  We also find
a phase of $P2_12_12_1$ symmetry that is marginally the most stable at
pressures around 144 GPa, see Fig.\
\ref{fig:Enthalpies_MgCO3_higher_pressure}.

\section{Structures and bonding}

The carbon atoms in the calcite, aragonite, ``post aragonite'', and
our {CaCO$_3$-$P2_1/c$-$l$} and {CaCO$_3$-$Pnma$-$h$} structures
contain threefold coordinated carbon atoms, as does the magnesite
phase of MgCO$_3$.  These structures contain triangular CO$_3^{2-}$
ions with $sp^2$ bonding.  In aragonite and ``post aragonite'' the
CO$_3^{2-}$ ions are coplanar, but in our {CaCO$_3$-$Pnma$-$h$}
structure they are somewhat tilted, while in {CaCO$_3$-$P2_1/c$-$l$}
they are tilted at approximately 90$^\circ$ to one another, see Fig.\
\ref{fig:structures_CaCO3_lower_pressure}.  More details of the
structures are given in the Supplemental Material \cite{Supplemental}.

\begin{figure}
  \centering
  \includegraphics[width=0.475\textwidth,clip]{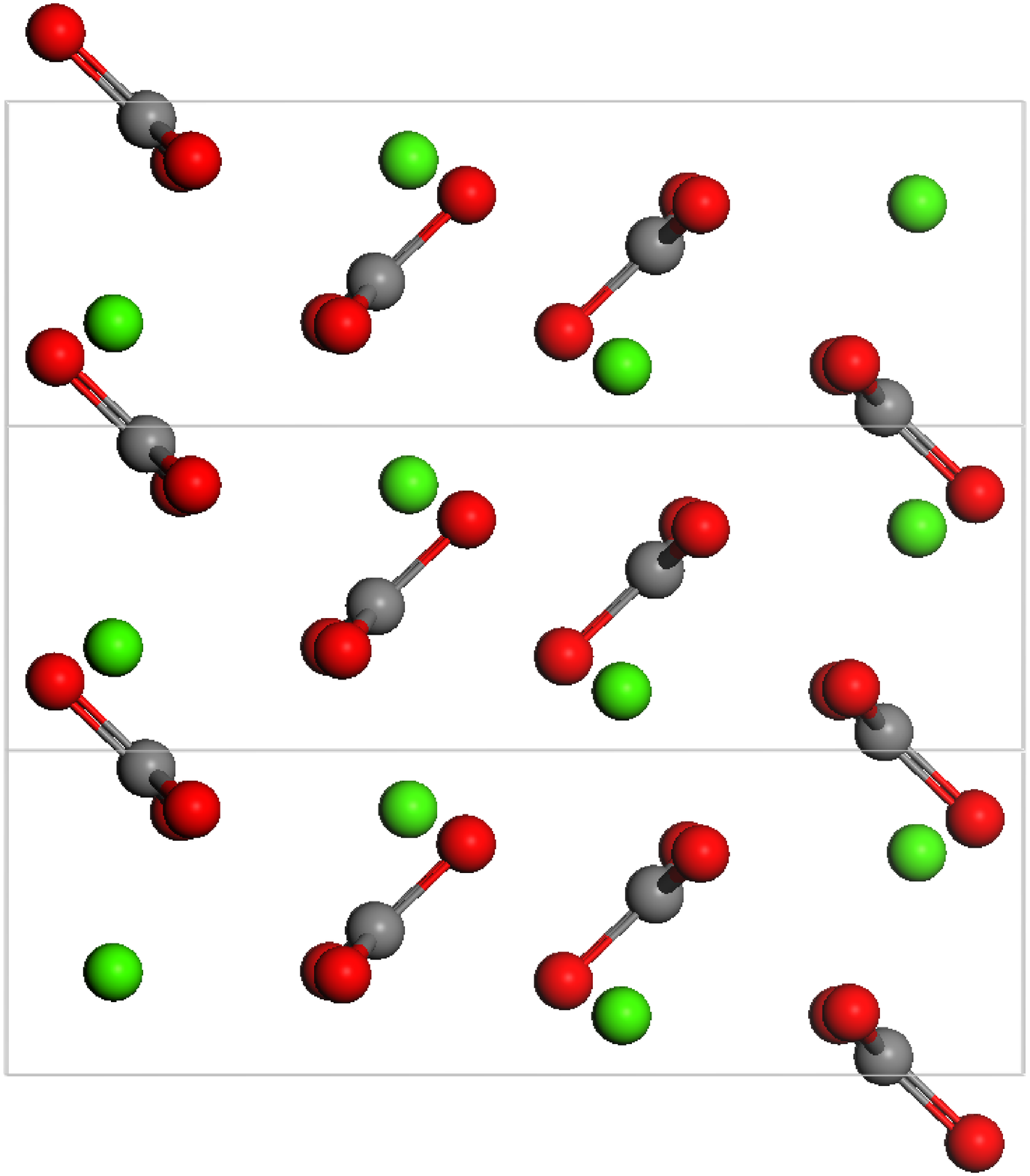}\\
  \includegraphics[width=0.37\textwidth,clip]{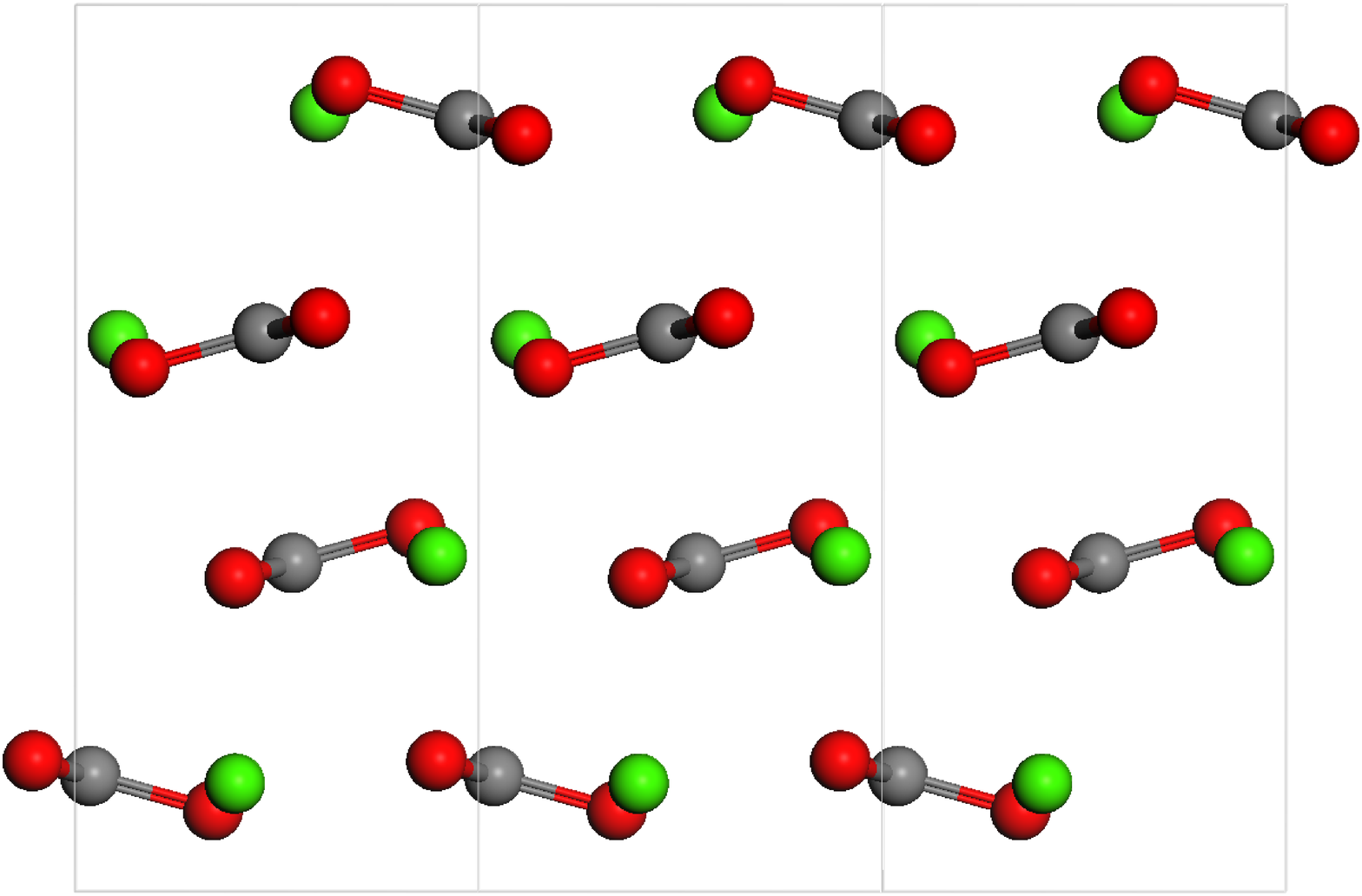}\\ \vspace{0.375cm}
  \includegraphics[width=0.3\textwidth,clip]{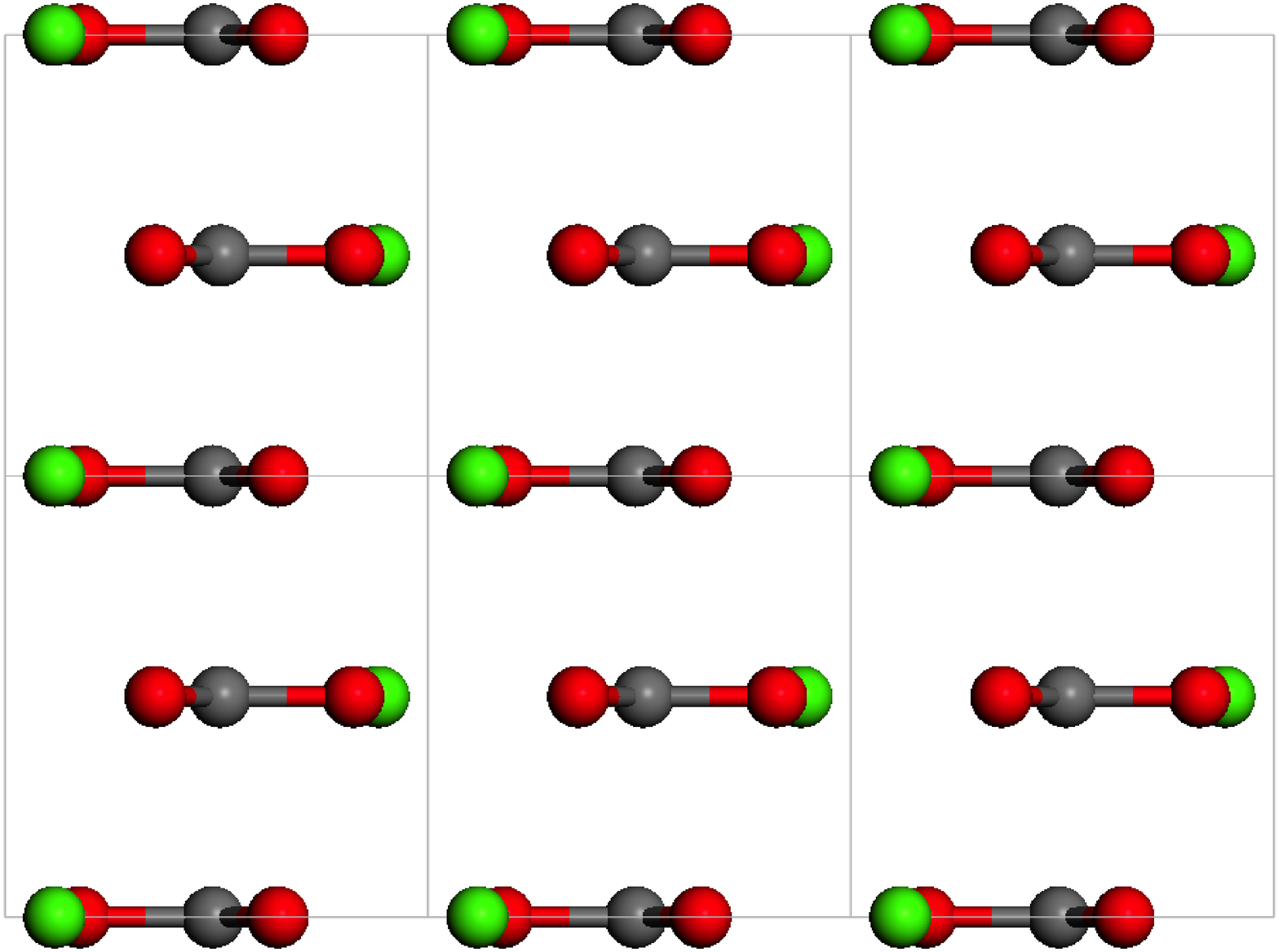}
  \caption{(Color online) The {CaCO$_3$-$P2_1/c$-$l$} (top),
    {CaCO$_3$-$Pnma$-$h$} (middle), and ``post aragonite'' (bottom)
    structures of CaCO$_3$ at 40 GPa.  The Ca atoms are in green, the
    carbon in grey, and the oxygen in red.}
  \label{fig:structures_CaCO3_lower_pressure}
\end{figure}

\begin{figure}
  \centering
  \includegraphics[width=0.35\textwidth,clip]{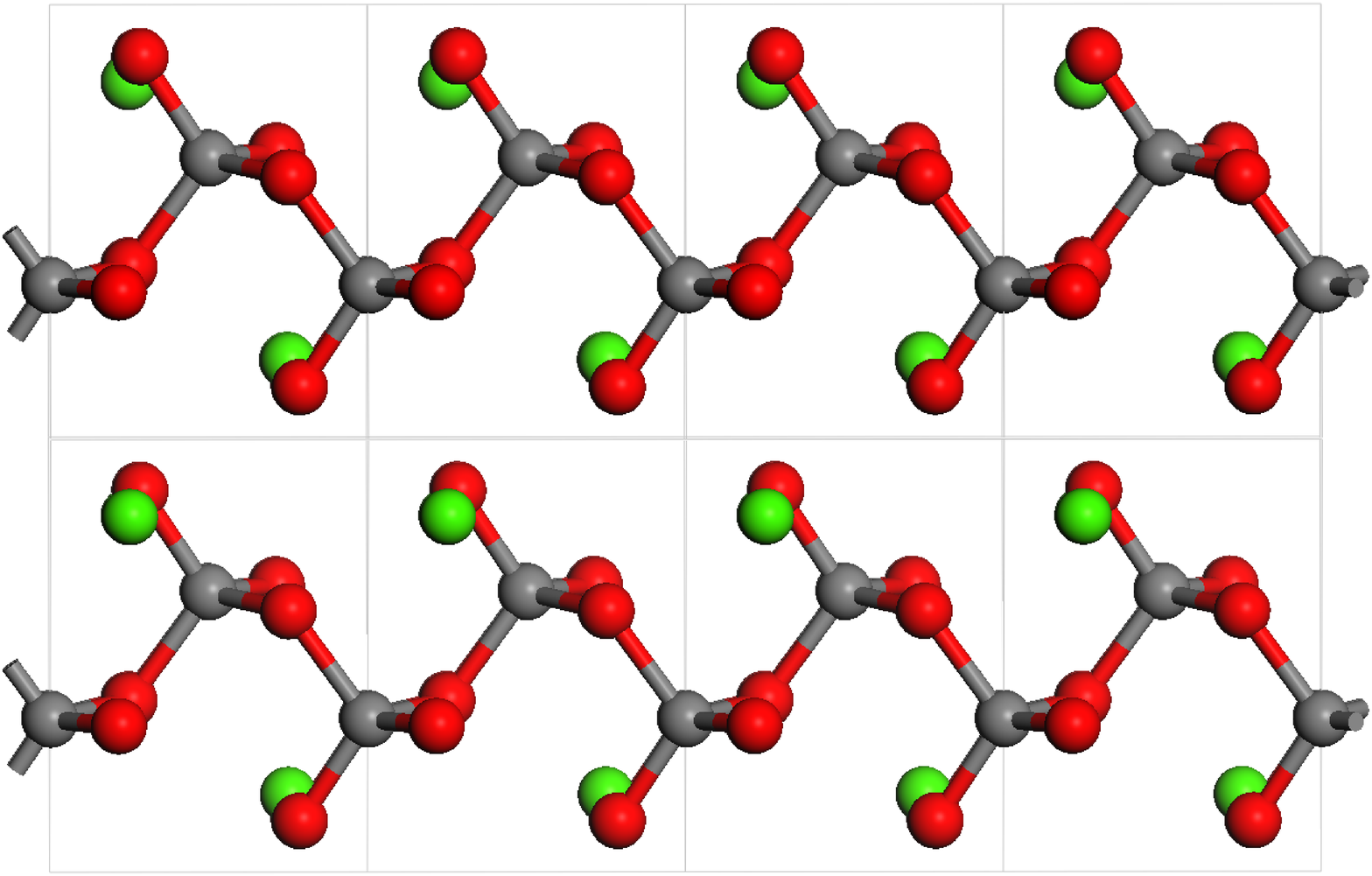}
  \includegraphics[width=0.35\textwidth,clip]{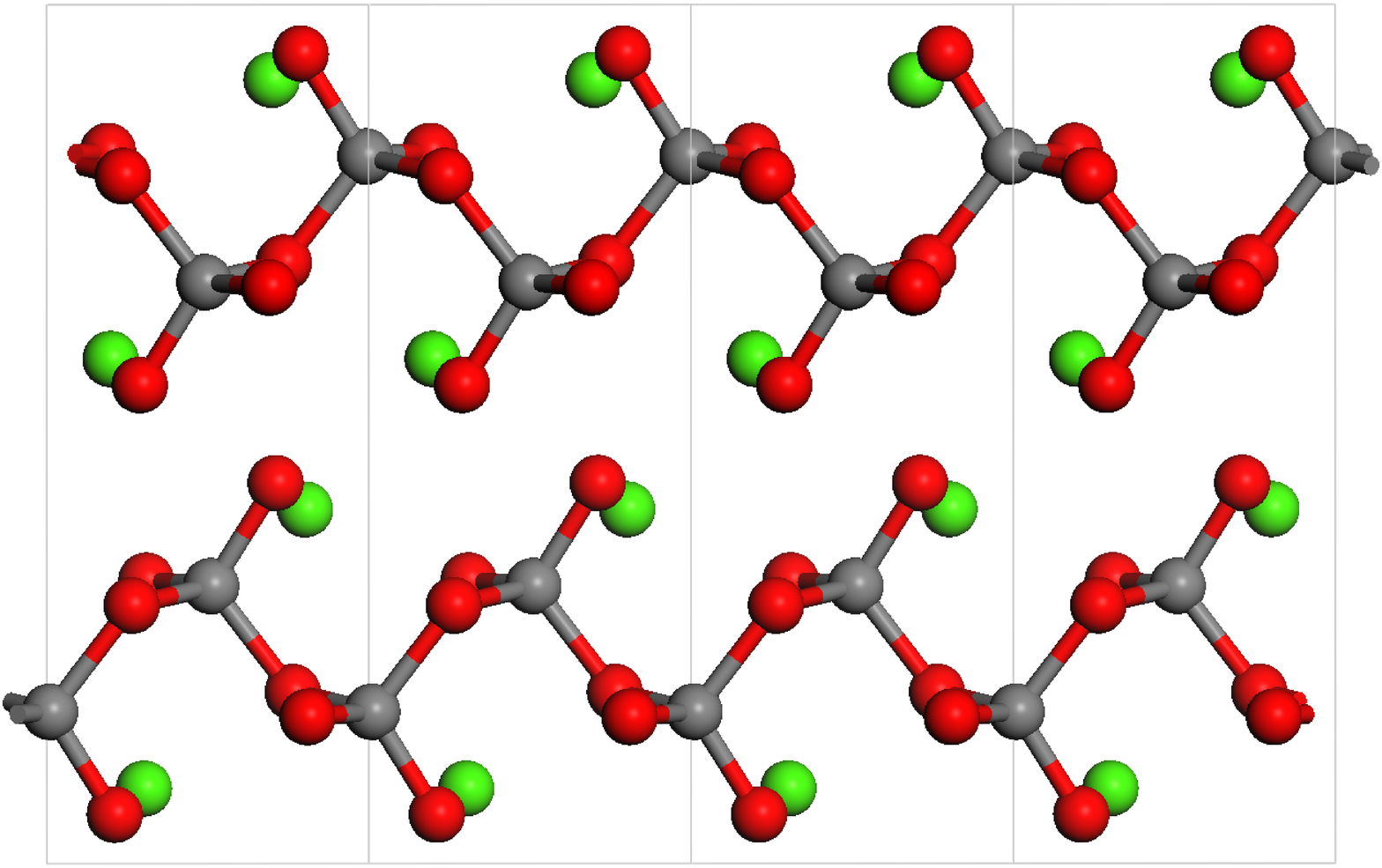}
  \caption{(Color online) The $C222_1$ (top) and {CaCO$_3$-$P2_1/c$-$h$}
    pyroxene-type (bottom) structures of CaCO$_3$ at 60 GPa. The Ca
    atoms are in green, carbon in grey, and oxygen in red.}
  \label{fig:structures_CaCO3_higher_pressure}
\end{figure}

The high-pressure {CaCO$_3$-$P2_1/c$-$h$} and $C222_1$ structures
contain fourfold coordinated carbon atoms and are of the pyroxene
type.  {CaCO$_3$-$P2_1/c$-$h$} and $C222_1$ possess very similar
calcium lattices but the packing of the pyroxene chains is different,
as can be seen in Fig.\ \ref{fig:structures_CaCO3_higher_pressure}.
In $C222_1$ each of the chains is orientated in the same manner, but
{CaCO$_3$-$P2_1/c$-$h$} alternate chains run in the reverse direction,
see Fig.\ \ref{fig:structures_CaCO3_higher_pressure}, and consequently
the unit cell of {CaCO$_3$-$P2_1/c$-$h$} contains four f.u., whereas
$C222_1$ contains two.  When viewed along the axis of the chains, the
{CaCO$_3$-$P2_1/c$-$h$} and $C222_1$ structures appear almost
identical.  {CaCO$_3$-$P2_1/c$-$h$} and $C222_1$ have very similar
volumes at high pressures, with $C222_1$ being slightly denser, which
leads to almost parallel enthalpy-pressure relations, see Fig.\
\ref{fig:enthalpy_CaCO3}.  The lower enthalpy of
{CaCO$_3$-$P2_1/c$-$h$} must therefore arise from more favorable
electrostatic interactions between the pyroxene chains.

\subsection{High-pressure X-ray data for C\lowercase{a}CO$_3$}

\begin{figure}
  \centering
 \includegraphics[width=0.4\textwidth,clip]{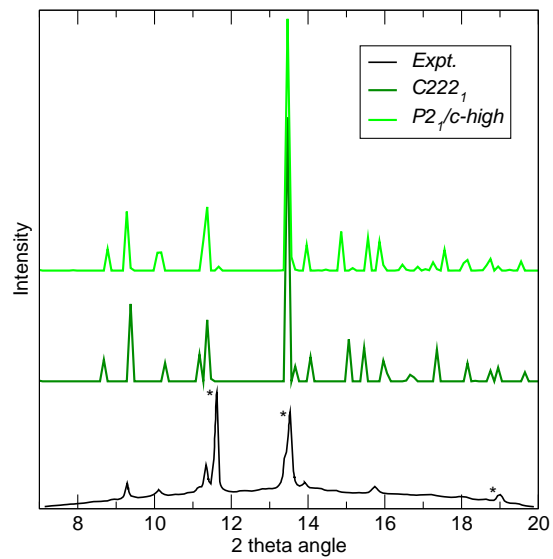}
 \caption{(Color online) X-ray diffraction patterns of the $C222_1$
   \cite{Oganov_2006} and {CaCO$_3$-$P2_1/c$-$h$} phases of CaCO$_3$,
   compared with experimental data from Fig.\ 1(b) of Ref.\
   \onlinecite{Ono_2007}.  Data at 182 GPa are reported, with an
   incident wavelength of 0.415 \AA.  The stars indicate that the peak
   immediately to the right arises from platinum.}
  \label{fig:XRD_CaCO3}
\end{figure}

Ono \textit{et al.}\ \cite{Ono_2007} performed laser-heated diamond
anvil cell experiments on CaCO$_3$ at 182 GPa.  X-ray diffraction data
for the $C222_1$ \cite{Oganov_2006} and {CaCO$_3$-$P2_1/c$-$h$}
structures are compared in Fig.\ \ref{fig:XRD_CaCO3} with the
experimental data from Fig.\ 1 of Ref.\ \onlinecite{Ono_2007}.  Note
the appearance of three peaks marked with stars in the experimental
data that arise from the platinum used to enhance heat absorption
during the laser heating and as a pressure calibrant.  The
experimental data is not of very high resolution.  The diffraction
patterns of the theoretical $C222_1$ and {CaCO$_3$-$P2_1/c$-$h$}
structures share many common features.  There are also clear
similarities between the theoretical and experimental X-ray data, but
the experimental data is of insufficient resolution to allow the
structure to be determined unambiguously.  We suggest that our
CaCO$_3$-$P2_1/c$-$h$ structure is the best available candidate for
the observed high pressure phase because it has a much lower enthalpy
than $C222_1$.

\section{Chemical reactions in Earth's mantle}

We have investigated possible chemical reactions involving the mantle
materials CaCO$_3$, MgCO$_3$, CO$_2$, MgSiO$_3$, CaSiO$_3$, SiO$_2$,
CaO and MgO, following the approach of Oganov \textit{et al.}\
\cite{Oganov_2008}.  The most stable structures of each compound at
the relevant pressures are used, as provided by DFT studies.  We use
the $Pa\bar{3}$, $P4_2/mnm$, and $I\bar{4}2d$ structures of CO$_2$
\cite{Ma_CO2_2013}, the stishovite, CaCl$_2$ and pyrite structures of
SiO$_2$ \cite{Karki_silica_1997}, the rocksalt structure of MgO, the
orthorhombic structure of perovskite CaSiO$_3$ and the perovskite and
post-perovskite structures of MgSiO$_3$ \cite{Murakami_2004,
  Oganov_2004, Tsuchiya_2004}.

\begin{figure}[h!]
  \centering
  \includegraphics[width=0.4\textwidth,clip]{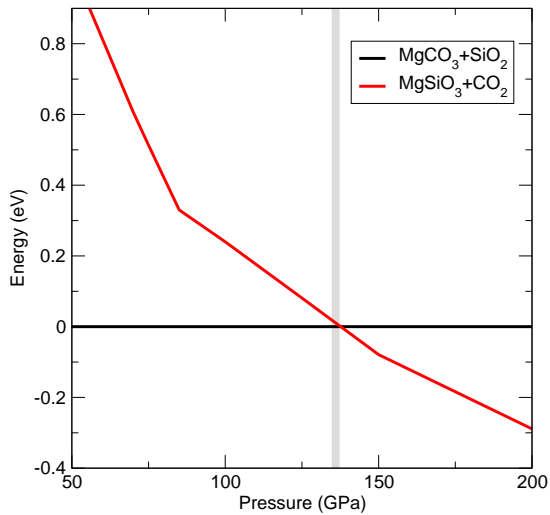}
  \caption{(Color online).  The relative stabilities per f.u.\ as a
    function of pressure of MgCO$_3$+SiO$_2$ and MgSiO$_3$+CO$_2$.
    The vertical gray line indicates the pressure at the base of the
    mantle (136 GPa).  In this and the following figures, the kinks
    arise from phase transitions. }
  \label{fig:Enthalpy_MgCO3+SiO2-MgSiO3+CO2}
\end{figure}

\begin{figure}[h!]
  \centering
  \includegraphics[width=0.4\textwidth,clip]{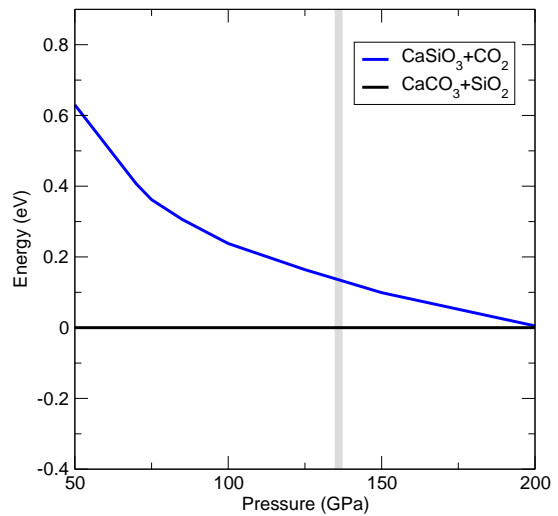}
  \caption{(Color online).  The relative stabilities per f.u.\ as a
    function of pressure of CaSiO$_3$+CO$_2$ and CaCO$_3$+SiO$_2$.
    The vertical gray line indicates the pressure at the base of the
    mantle (136 GPa). }
  \label{fig:Enthalpy_CaSiO3+CO2-CaCO3+SiO2}
\end{figure}

Decomposition of CaCO$_3$ and MgCO$_3$ into the alkaline earth oxides
plus CO$_2$ is found to be unfavorable.  Under conditions of excess
SiO$_2$, the reaction
\begin{eqnarray}
{\rm MgCO}_3 + {\rm SiO}_2 & \rightarrow & {\rm MgSiO}_3 + {\rm CO}_2
\end{eqnarray}
is found to be energetically unfavorable up to 138 GPa, which is just
above the pressure at the mantle-core boundary, see Fig.\
\ref{fig:Enthalpy_MgCO3+SiO2-MgSiO3+CO2}.  We find that the reaction
\begin{eqnarray}
{\rm CaCO}_3 + {\rm SiO}_2 & \rightarrow & {\rm CaSiO}_3 + {\rm CO}_2
\end{eqnarray}
does not occur below 200 GPa, see Fig.\
\ref{fig:Enthalpy_CaSiO3+CO2-CaCO3+SiO2}, which is much higher than
the value of 135 GPa reported in Ref.\ \onlinecite{Oganov_2008}.  We
conclude that both MgCO$_3$ and CaCO$_3$ are stable within the Earth's
mantle under conditions of excess SiO$_2$.  These results suggest that
free CO$_2$ does not occur as an equilibrium phase within the Earth's
mantle.

\begin{figure}[h!]
  \centering
  \includegraphics[width=0.45\textwidth,clip]{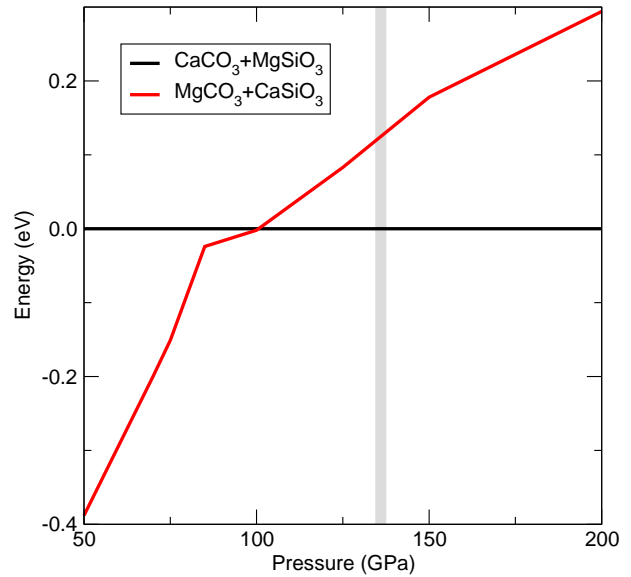}
  \caption{(Color online) Enthalpy per f.u.\ of CaCO$_3$+MgSiO$_3$
    compared with that of CaSiO$_3$+MgCO$_3$.  Below 100 GPa we find
    that CaSiO$_3$+MgCO$_3$ is the most stable, while above 100 GPa
    CaCO$_3$+MgSiO$_3$ is the most stable. }
  \label{fig:Enthalpy_CaCO3+MgSiO3-CaSiO3+MgCO3}
\end{figure}

MgCO$_3$ has generally been believed to be the dominant carbonate
throughout the Earth's mantle.  This assumption can be tested when
excess MgO is present by determining the relative stability of
CaCO$_3$+MgO and MgCO$_3$+CaO.  We find that CaCO$_3$+MgO is the more
stable up to pressures of about 200 GPa, so that CaCO$_3$ is the
stable carbonate under these conditions.  In the case of excess
MgSiO$_3$ we consider the reaction
\begin{eqnarray} 
{\rm CaCO}_3 + {\rm MgSiO}_3 & \rightarrow & {\rm CaSiO}_3 + {\rm MgCO}_3,
\end{eqnarray}
finding that CaCO$_3$ is more stable than MgCO$_3$ from 100 GPa up to
pressures well above those of 136 GPa found at the mantle-core
boundary, see Fig.\ \ref{fig:Enthalpy_CaCO3+MgSiO3-CaSiO3+MgCO3}.

\section{Conclusions}

In conclusion, we have searched for structures of CaCO$_3$ and
MgCO$_3$ at mantle pressures using AIRSS
\cite{Airss_review,pickard_silane}.  We have found a
{CaCO$_3$-$P2_1/c$-$l$} structure with $sp^2$ bonded carbon atoms that
is predicted to be stable within the range 32--48 GPa.  We have also
found a high pressure {CaCO$_3$-$P2_1/c$-$h$} structure with $sp^3$
bonded carbon atoms that is about 0.18 eV per f.u.\ more stable than
the $C222_1$ phase proposed by Oganov \textit{et al.}\
\cite{Oganov_2006}.  Both the {CaCO$_3$-$P2_1/c$-$h$} and $C222_1$
structures are compatible with the available X-ray diffraction data
\cite{Ono_2007}.  However, {CaCO$_3$-$P2_1/c$-$h$} is the most stable
structure from 67 GPa to pressures well above those encountered within
the Earth's lower mantle ($\leq$ 136 GPa).
Our AIRSS calculations suggest a previously unknown phase of MgCO$_3$
of $P\bar{1}$ symmetry that is predicted to be thermodynamically
stable in the pressure range 85--101 GPa.
Our results suggest that CO$_2$ is not a thermodynamically stable
compound under deep mantle conditions.
Under conditions of excess MgSiO$_3$ we find that CaCO$_3$ is more
stable than MgCO$_3$ above 100 GPa.  This result arises directly from
our discovery of the highly stable {CaCO$_3$-$P2_1/c$-$h$} phase.  The
results of our study change our understanding of the carbon cycle in
the lower part of the mantle and may have important consequences for
geodynamics
\cite{Javoy_carbon_geodynamic_cycle,Marty_1987_geodynamics,Dobretsov_2012_geodynamics}
and geochemistry
\cite{Schrag_2013_geochemistry,Deines_geochem_review}.

\begin{acknowledgments}

  We acknowledge financial support from the Engineering and Physical
  Sciences Research Council (EPSRC) of the United Kingdom.

\end{acknowledgments}

\end{document}